\documentclass{ws-ijmpe}

\begin{document}

\markboth{P. Olbratowski et al.}{HFB calculations for nuclei with
tetrahedral deformation}

%
\catchline{}{}{}{}{}
%

\title{HARTREE-FOCK-BOGOLYUBOV CALCULATIONS FOR NUCLEI WITH TETRAHEDRAL DEFORMATION}

\author{P. OLBRATOWSKI and J. DOBACZEWSKI}

\address{Institute of Theoretical Physics, Warsaw University, \\
ul. Ho\.za 69, 00-681, Warsaw, Poland}

\author{P. POWA\L{}OWSKI and M. SADZIAK}

\address{Department of Physics, Warsaw University, \\
ul. Ho\.za 69, 00-681, Warsaw, Poland}

\author{K. ZBERECKI}

\address{Faculty of Physics, Warsaw University of Technology, \\
ul. Koszykowa 75, 00-662 Warsaw, Poland}

\maketitle

\begin{history}
\received{(received date)}
\revised{(revised date)}
\end{history}

\begin{abstract}
Hartree-Fock-Bogolyubov solutions corresponding to the tetrahedral
deformation are found in six tetrahedrally doubly-magic nuclei.
Values of the $\beta_{32}$ deformation, depths of the tetrahedral
minima, and their energies relative to the co-existing quadrupole
minima are determined for several versions of the Skyrme force.
Reduction of the tetrahedral deformation energies by pairing
correlations is quantitatively analysed. In light nuclei, shallow
tetrahedral minima are found to be the lowest in energy, while in
heavy nuclei, the minima are deeper but appear at a few MeV of
excitation.
\end{abstract}

\section{Introduction}

It is known that an increased nuclear binging is caused by the
presence of large energy gaps in the single-particle (s.p.) nuclear
spectra. The large gaps result in a decreased average density of
s.p.\ levels and influence binding energies through the so-called
shell effects. These effects can be further enhanced by high
degeneracies of the s.p.\ levels above and/or below the energy gaps,
which results in even larger fluctuations of the average level
densities. Such degeneracies, in turn, are consequences of the
conservation of certain symmetries in the s.p.\ Hamiltonian. Ordinary
doubly-magic nuclei, for example, are spherically symmetric, i.e.\
characterized by degenarcies corresponding to the rotational group
$O(3)$, and indeed the most strongly bound. Apart from the group of
rotations, there exist only two other relevant symmetry groups whose
conservation leads to s.p.\ degeneracies higher than the two-fold
Kramers degeneracy. One of them is the point group, $T_d$, of the
regular tetrahedron, which yields two-fold and four-fold degenerate
s.p.\ levels. On this basis, Li and Dudek\cite{Li94a} suggested in
1994 that stable nuclear shapes characterized by the tetrahedral
symmetry may exist in Nature.

The lowest-rank multipole deformation which does not violate the
$T_d$ symmetry is $\beta_{32}$\cite{Dud03a}. It represents a shape
of a regular tetrahedron with "rounded edges and corners", and is
usually called tetrahedral deformation. By using the deformed
Woods-Saxon potential, several authors\cite{Li94a,Dud03a,Dud02a}
examined the s.p.\ energies in function of $\beta_{32}$, and found
that, indeed, large energy gaps, sometimes larger than the spherical
ones, open up at neutron/proton numbers of N/Z=16, 20, 32, 40, 56-58,
70, 90-94, 100, 112, 126 or 136. They are sometimes referred to as
tetrahedral magic numbers. In the vicinity of the tetrahedrally
doubly-magic nuclei defined in this way, Strutinsky shell-correction
calculations were performed\cite{Li94a,Dud02a,Sch04a,Sch04b}, and
energy minima corresponding to the tetrahedral deformation were found
in even-even $^{80}$Zr, $^{106-112}$Zr, $^{160}$Yb, $^{222}$Rn, and
$^{242}$Fm. Similarly, the Hartree-Fock+BCS (HF+BCS) calculations
\cite{Tak98a}, found tetrahedral solutions in $^{80}$Zr,
and Hartree-Fock-Bogolyubov (HFB) tetrahedral solutions in $^{80}$Zr and $^{106-112}$Zr were reported in Refs.\,\cite{Yam01a} and \cite{Sch04a,Sch04b}, respectively.

The present paper reports on the first systematic study of the
tetrahedral deformation in various regions of the nuclear chart,
carried out by means of self-consistent methods. We focus our study on
properties of the tetrahedral minima, mainly their energies and
deformations, and analyze their dependence on the Skyrme force
parameterizations.

\section{HFB calculations}

The HFB method was used. Four parameter
sets of the Skyrme interaction were taken in the particle-hole
channel: SLy4\cite{Cha97a}, SkM*\cite{Bar82a}, SkP\cite{Dob84a},
and SIII\cite{Bei75a}. For the description of pairing, procedures of
Ref.\,\cite{Dob02a} were followed. The density-dependent delta
interaction in the form of
\begin{equation}
V(\vec{r}_1,\vec{r}_2)=V_0\left(1-\frac{\rho(\vec{r}_1)}{2\rho_0}\right)\delta(\vec{r}_1-\vec{r}_2)
\end{equation}
was employed, with the saturation density, $\rho_0$=0.16\,fm$^{-3}$,
and strengths of $V_0$=$-$285.88, $-$233.94, $-$213.71, and
$-$249.04\,MeV\,fm$^3$ for SLy4, SkM*, SkP, and SIII, respectively.
The densities were constructed out of quasi-particle states with
equivalent-spectrum energies\cite{Dob84a} up to 60\,MeV. In order to
study effects of pairing, the HF calculations were also performed for
comparison. Reflection symmetries in two or three perpendicular planes
were imposed, correspondingly, when looking for the tetrahedral and
quadrupole solutions. The calculations were carried out by using the
code HFODD (v2.11k)\cite{Dob00a,Dob04a,Dob05a}, which expands the
quasi-particle wave-functions onto the Harmonic-Oscillator basis.
Bases of 14 and 16 spherical shells were taken for the Zr and heavier
elements, respectively.

Six nuclei, doubly-magic with respect to the tetrahedral magic
numbers, were examined: $^{80}_{40}$Zr$_{40}$, $^{98}_{40}$Zr$_{58}$,
$^{110}_{40}$Zr$_{70}$, $^{126}_{56}$Ba$_{70}$,
$^{160}_{70}$Yb$_{90}$, and $^{226}_{90}$Th$_{136}$. In all of them,
the HF and HFB energy minima corresponding to the tetrahedral shapes were
found with all the examined forces, apart from a few exceptions. The found
solutions are characterized by the $\beta_{32}$ deformations ranging
from 0.08 to 0.26, and admixtures of $\beta_{40}$ and $\beta_{44}$
deformations in proportions that preserve the tetrahedral
symmetry\cite{Dud03a}, and with values of $\beta_{40}$ ranging from
about 0.01 to 0.07. Deformations of higher multipolarities were found
to be negligibly small. As expected for the tetrahedral symmetry, the
HF s.p.\ and HFB quasi-particle spectra are composed of two-fold and
four-fold degenerate levels. In the six nuclei in question,
spherical, oblate, prolate, and triaxial solutions were also found,
depending on the nucleus, as discussed below.

\begin{figure}[thp]
\centerline{\psfig{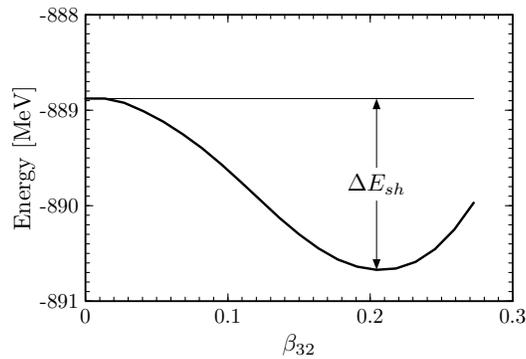}}
\vspace*{8pt}
\caption{Total energy in function of the tetrahedral deformation
$\beta_{32}$ in $^{110}$Zr, obtained from the HFB calculations with the
SIII force. $\Delta E_{sh}$ denotes the energy difference between the
spherical point and the tetrahedral minimum.}
\label{fig1}
\end{figure}

Three quantities that characterize the obtained tetrahedral solutions
will be examined: the energy difference, $\Delta E_{hq}$, between the
tetrahedral ($h$) and lowest quadrupole ($q$) minima, the energy
difference, $\Delta E_{sh}$, between the spherical point ($s$) and
tetrahedral minimum, and the deformation $\beta_{32}$. $\Delta
E_{hq}$ gives an idea of the excitation energy of the tetrahedral
states above the ground state. $\Delta E_{sh}$ is important for the
following reason. Both from the previous self-consistent studies in
$^{80}$Zr\cite{Tak98a,Yam01a}, as well as from our
preliminary results for $^{80,98,110}$Zr, it seems that, at least in
the Zr isotopes, there is no energy barrier between the spherical and
tetrahedral solutions. Energy in function of $\beta_{32}$ looks
rather like the dependence shown in Fig.\ \ref{fig1}, obtained from the HFB
calculations in $^{110}$Zr with the SIII force. One can see that
$\Delta E_{sh}$ measures the depth of the tetrahedral minimum against
changes in $\beta_{32}$, and thus provides information on whether a
stable tetrahedral deformation or rather tetrahedral vibrations about
the spherical shape should be expected.

\begin{figure}[thp]
\centerline{\psfig{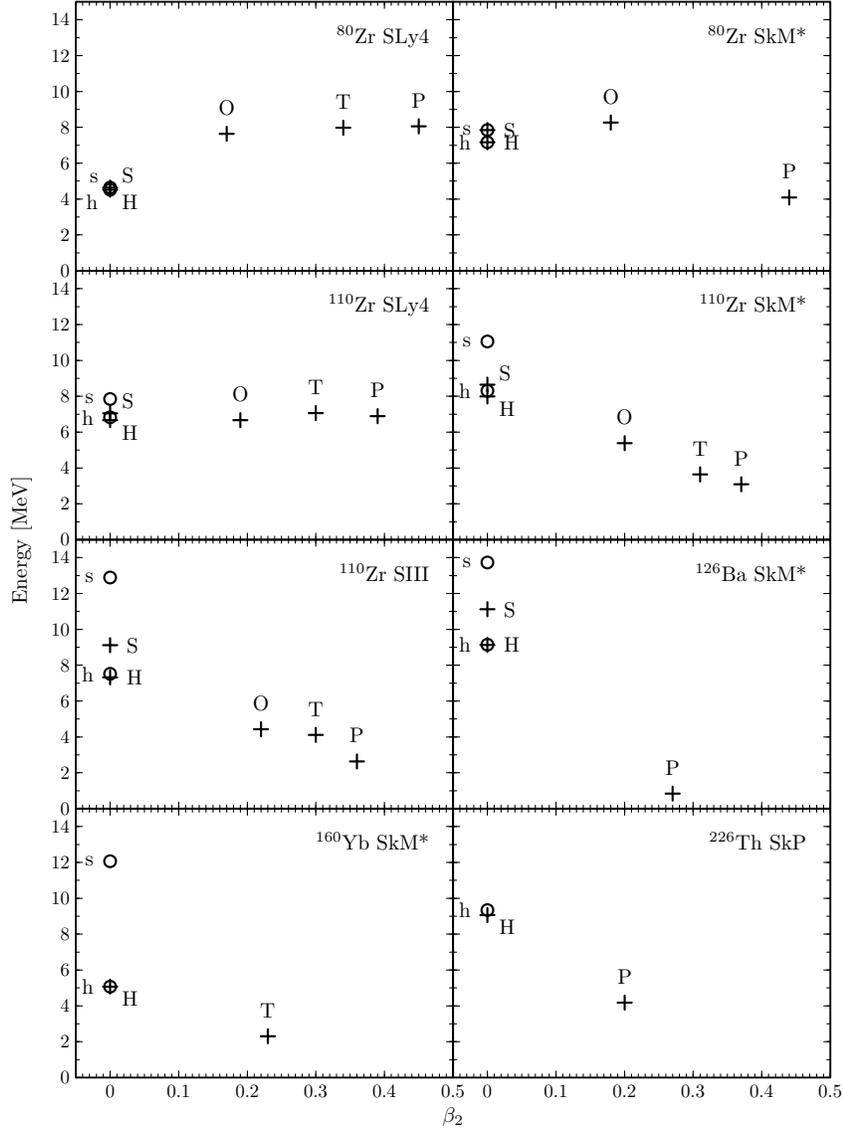}}
\vspace*{8pt}
\caption{The HF (circles, lower-case labels) and HFB (plus symbols,
upper-case labels) energy minima marked on the $\beta_2$-$E$ plane
for selected nuclei and forces, as specified in each panel. The
tetrahedral ($h$, $H$), spherical ($s$, $S$), oblate ($O$), prolate
($P$), and triaxial ($T$) solutions are shown.}
\label{fig2}
\end{figure}

Figure \ref{fig2} shows the energy minima for selected nuclei and forces as
points on the $\beta_2$-$E$ plane. In each panel, all the found HFB
solutions (plus symbols, upper-case labels) are shown, while the HF
solutions (circles, lower-case labels), are given only for the
tetrahedral and spherical cases. The labels denote the tetrahedral
($h$, $H$), spherical ($s$, $S$), oblate ($O$), prolate ($P$), and
triaxial ($T$) solutions.

One of the principal observations resuilting from our calculations
is that various Skyrme forces may give
significantly different energetical positions of the tetrahedral
solutions with respect to the quadrupole minima. This is most
pronounced for the SLy4 and SkM* results in $^{80}$Zr (two upmost
panels in Fig.\ \ref{fig2}), which give the tetrahedral minima of about 3\,MeV
below and above the lowest quadrupole state, respectively. The extremal
values of $\Delta E_{hq}$ predicted by various forces for each
nucleus studied here are collected in Table 1. They do not depend much on whether
pairing is included or not. The differences between the maximum and minimum
values are of the order of a few MeV. One can see, nevertheless, that $\Delta E_{hq}$ is
lowest in Zr isotopes, where some forces even predict the tetrahedral
solution to be the ground state. Values of $\Delta E_{hq}$ are
particularly large, not smaller than 7\,MeV, in $^{126}$Ba and rather
moderate, even about 3\,MeV, in $^{160}$Yb and $^{226}$Th.

\newcommand{\myrule}  {\rule{0mm}{3.3mm}}
\begin{table}[pt]
\tbl{Minimum (min) and maximum (max) values of $\Delta E_{hq}$ [MeV],
$\Delta E_{sh}$ [MeV], and $\beta_{32}$ from among the results
obtained with various Skyrme forces. Results are shown for each nucleus and
approximation (HF or HFB) studied here.}
{\begin{tabular}{r|cc|cccc|cccc}
\hline
\myrule    & \multicolumn{2}{c|}{$\Delta E_{hq}$} & \multicolumn{4}{c|}{$\Delta E_{sh}$} & \multicolumn{4}{c}{$\beta_{32}$} \\
\cline{2-11}
\myrule    & \multicolumn{2}{c|}{HFB} & \multicolumn{2}{c}{HF} & \multicolumn{2}{c|}{HFB} & \multicolumn{2}{c}{HF} & \multicolumn{2}{c}{HFB} \\
Nucleus    & min  & max & min  & max & min  & max & min & max & min & max \\
\hline\myrule
 $^{80}$Zr & -3   &  3  & 0.1  & 2   & 0.1  & 2   & 0.11 & 0.20 & 0.11 & 0.20 \\
 $^{98}$Zr & -1.5 &  2  & 0.3  & 0.7 & 0.04 & 0.5 & 0.14 & 0.16 & 0.09 & 0.20 \\
$^{110}$Zr & -0.4 &  5  & 0.07 & 5   & 0.4  & 2   & 0.08 & 0.23 & 0.14 & 0.21 \\
$^{126}$Ba &  7   & 11  & 3    & 5   & 2    & 2   & 0.17 & 0.26 & 0.22 & 0.22 \\
$^{160}$Yb &  3   &  7  & 3    & 7   &      &     & 0.19 & 0.26 & 0.23 & 0.24 \\
$^{226}$Th &  3   &  8  &      &     &      &     & 0.18 & 0.24 & 0.15 & 0.22 \\
\hline
\end{tabular}}
\end{table}

Further important point is that the depths of the tetrahedral minima,
$\Delta E_{sh}$, are reduced by pairing. This can be seen from the
comparison of the HF and HFB results for $^{110}$Zr and $^{126}$Ba
(four central panels in Fig.\ \ref{fig2}). The reductions may be as significant
as from 3 to 1\,MeV in $^{110}$Zr with SkM*, while for SkP the
inclusion of pairing suppresses the tetrahedral minimum in $^{110}$Zr altogether.
The decrease in $\Delta E_{sh}$ is mainly due to the lowering of the
energy at the spherical point, i.e., the pairing influences the
spherical state more than the tetrahedral one. This is so because the
s.p.\ energy gaps at the Fermi level are bigger in the latter case,
as already discussed in the Introduction. In $^{80}$Zr with SIII, for
instance, pairing vanishes at the tetrahedral minimum, and remains
non-zero at the spherical point. Predictions concerning the
destructive role of pairing strongly depend on the details of the
method, as well. The HF+BCS \cite{Tak98a} and HFB\cite{Yam01a}
calculations for $^{80}$Zr, both using the SIII force, yielded
$\Delta E_{sh}$ of 0.7\,MeV and several tens of keV, respectively.
The corresponding result of our calculations is about 2\,MeV.

In the current analysis, we also obtain differences in predictions of
various Skyrme forces as to the values of $\Delta E_{sh}$, both with
and without pairing. In the HFB results for $^{110}$Zr, for example,
$\Delta E_{sh}$ varies from about 0.4 for SLy4 to 2\,MeV for SIII,
not counting SkP. The HF and HFB results for other studied nuclei are
summarized in Table 1. In $^{226}$Th, no spherical solutions, and in
$^{160}$Yb no spherical HFB solutions were found, so that the
corresponding values of $\Delta E_{sh}$ could not be calculated. In
$^{126}$Ba, the tetrahedral HFB solution was obtained with only one
force, SkM*. However, the HF results exhibit a clear trend that
$\Delta E_{sh}$ increases with the mass number. It can be as small as
a few tens of keV in Zr isotopes, and as large as 7\,MeV for
$^{160}$Yb with SkM*. The problem of stability of the
tetrahedral minima can be even more complicated because of a possible
softness of the nuclei in question against the octupole deformations
other than $\beta_{32}$\cite{Tak98a,Yam01a}.

The four Skyrme forces used in our study also give somewhat different
values of the $\beta_{32}$ deformation, see Table 1. The inclusion
of pairing slightly reduces these values, along with $\Delta E_{sh}$.
Heavier isotopes have larger values of $\beta_{32}$.

\section{Summary}

Hartree-Fock-Bogolyubov solutions corresponding to the tetrahedral
deformation were found in six tetrahedrally doubly-magic nuclei:
$^{80}$Zr, $^{98}$Zr, $^{110}$Zr, $^{126}$Ba, $^{160}$Yb, and
$^{226}$Th. Results with four Skyrme forces, SLy4, SkM*, SIII, and
SkP, sometimes significantly differ in the values of $\beta_{32}$,
depths of the tetrahedral minima, and their energies with respect to
the co-existing quadrupole solutions. The inclusion of pairing
reduces the depths, or even suppresses the existence of the
tetrahedral minima. In Zr isotopes, the tetrahedral minima are rather shallow,
but some forces predict them as the lowest in energy. In $^{126}$Ba,
they are not lower than 7\,MeV above the quadrupole solutions, but in
$^{160}$Yb and $^{226}$Th that minimum distance is reduced to 3\,MeV.
Tetrahedral minima in $^{126}$Ba, $^{160}$Yb, and $^{226}$Th are
estimated to be deeper than in the Zr isotopes.

\section*{Acknowledgements}

This work was supported in part by the Polish Committee for
Scientific Research (KBN) under Contract No.~5~P03B~014~21 and by the
Foundation for Polish Science (FNP).

\end{document}